\documentclass[10pt,twocolumn,floatfix]{revtex4}
\usepackage{inputenc}
\usepackage{graphicx,epsfig,amssymb,multirow}

\begin{document}

\title{A study of the density functional methods on the photoabsorption of bodipy dyes}

\author{Hatice \"{U}nal$\,^1$, Deniz Gunceler$\,^2$, Ersen Mete$\,^{1,}$\footnote{Corresponding author: \indent e-mail:
emete@balikesir.edu.tr}}
\affiliation{$^1$Deparment of Physics, Bal{\i}kesir University, Bal{\i}kesir 10145, Turkey}
\affiliation{$^2$Deparment of Physics, Cornell University, Ithaca, NY 14853, USA}

\begin{abstract}
Tunability of the photoabsorption and directional charge injection 
characteristics of Bodipy-based dye molecules with different carbonyl 
groups make them promising candidates for photovoltaic applications. 
In order to study the effect of screening in the Coulomb interaction on the 
electronic and optical properties of two Bodipy derivatives, we have used 
linear response time-dependent and exact exchange hybrid density functional 
approaches. The effect of linear and non-linear solvation models on the 
electrochemical properties of the dyes has also been discussed.
\end{abstract}



\maketitle

\section{Introduction}

The dye-sensitized solar cells (DSSC) form an important line of research as 
renewable energy sources.\cite{Fujishima,Regan} As being simple, economical 
and environmentally friendly, the DSSC attracts attention as a good alternative
over expensive thin film solar cell technologies, The energy conversion 
efficiency-price ratio allows it to compete with the other energy generation 
technologies.  The strong demand for a reasonable efficiency drives 
researchers to improve and find better sensitizers. Bodipy-based dyes are 
promising in light harvesting with their modifiable structures by adding 
different functional groups.\cite{Ela,Kolemen} The stability and efficiency of 
a cell mostly depend on the characteristics of the sensitizer. Light-harvesting 
electronic spectra, adsorption, charge injection efficiency and durability of 
the dye molecule during the cycles are important features for DSSC applications. 
In addition, their purification and cost should also be taken into consideration. 
For this reason, instead of metal-driven dye molecules~\cite{Tachibana,Thompson}, 
organic sensitizers for DSSC applications have become the key issue of the many 
researchers.
 
\begin{figure*}[t!]
\epsfig{file=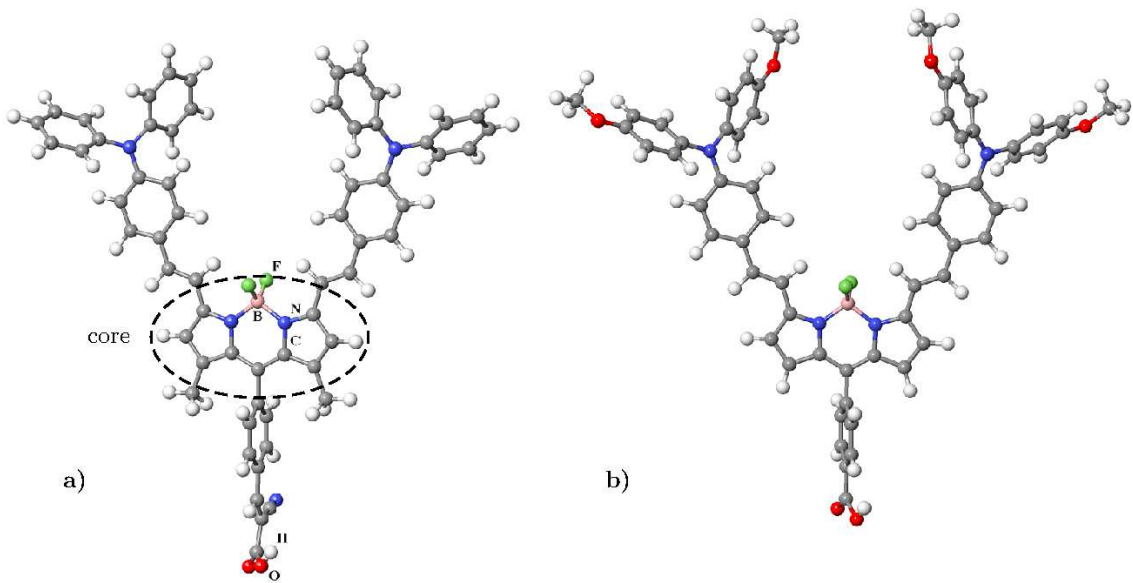,width=15cm}
\caption{Minimum energy geometries of Panchromatic Boradiazaindacene dye molecules 
a) Bodipy1 and b) Bodipy2.\label{fig1}}
\end{figure*}

Boron dipyrrins, which are widely known as Bodipy dyes, provide high incident 
photon current efficiency (IPCE) for DSSCs.  The adjustable structures of Bodipy 
dyes allow them to absorb all frequencies within the visible range extending 
to near IR~\cite{Ela,Kolemen,Hattori,Loudet,Kumaresan}. The Bodipy compounds were 
reported as supporting the directional transfer of charge to excitation state. 
Such a mechanism might help reduce photo-generated electron-hole pair 
recombination and enhance charge injection efficiency. In this manner we consider 
two derivatives referred as Bodipy1 and Bodipy2 (Fig.~\ref{fig1}) which 
were synthesized by Erten-Ela~\textit{et al.} and Kolemen~\textit{et al.}, 
respectively~\cite{Ela,Kolemen}. The parent Bodipy dye condensed through 
phenyl substituents as donor groups and cyanoacetic acid derived 
electron-withdrawing anchor groups~\cite{Ela,Kolemen}.

We have studied the absorbance of the two Bodipy-derived molecules using 
time-dependent density functional perturbation theory~\cite{Runge,Gross,Casida,Marques,Malcioglu} 
as well as using the hybrid method fractionally incorporating the exact exchange 
interaction with and without its long-range part.\cite{Heyd1,Heyd2,Paier,pbe0} 
Our primary aim is to shed light on how these methods describe optical absorption 
spectra of such molecular structures with charge transfer (CT) excitations with
an emphasis on the charge localization effects due to screening of the Coulomb 
interaction. In order to make a reasonable comparison with available experimental 
results, we repeated our calculations including a recent solvation model~\cite{Gunceler} 
that also elucidates the effect of chloroform solution in which the dye absorbs light.

\section{Computational Details}

We performed first-principles total-energy calculations using ultrasoft 
pseudopotentials (USPPs)~\cite{Vanderbilt} in plane wave based DFT. 
The exchange-correlation contributions have been treated using 
Perdew--Burke--Ernzerhof (PBE)~\cite{pbe} gradient corrected 
functional as implemented in the Quantum Espresso 
package~\cite{Scandolo,Giannozzi,espresso}.

After having Kohn-Sham states and energies, optical properties were computed 
using turboTDDFT code~\cite{Malcioglu}, which is one of the modules of the 
Quantum Espresso code and implements the Liouville-Lanczos 
approach~\cite{Walker1,Rocca1} to TDDFT~\cite{Runge,Gross,Casida,Marques}. 
The response of the molecular systems to an external field is measured by 
dynamic molecular polarizability, 
\[\alpha_{ij}(\omega)=\textrm{Tr}(X_i\rho'_j(\omega))\]
where $X_i$ is the $i$-th component of the position operator, $\rho'_j(\omega)$ 
is the response density matrix of the system polarized along the $j$-th axis, 
oscillating at the frequency $\omega$.\cite{Malcioglu} The polarizability 
can be expressed as an off-diagonal matrix element of the resolvent 
of the Liouvillian super-operator and can be evaluated using the Lanczos 
algorithm, by means of the representation of density matrices transferred 
from time-independent density functional perturbation theory.
\cite{Malcioglu,Walker1,Rocca1,Walker2,Rocca2} The Liouville-Lanczos 
method implemented time-dependent density functional perturbation theory 
(TDDFPT) calculation provides spectroscopic features of molecular systems 
that is composed of hundreds of atoms. Since this approach  does not require 
the calculation of the individual Liouvillian eigenvalues and allows to reach 
the full optical spectrum directly, it needs a computational effort which is not 
so large than a standard DFT calculation.\cite{Malcioglu}

The next method that we used is the hybrid density functional approach 
which partially admixes exact Hartree-Fock (HF) and PBE exchange energies.  
One of the well known flavors is the PBE0 exchange-correlation (XC) 
functional defined as,
\[
E_{\tiny\textbf{XC}}^{\scriptsize\textrm{PBE0}}=
\frac{1}{4}E_{\tiny\textbf{X}}^{\scriptsize\textrm{HF}}+
\frac{3}{4}E_{\tiny\textbf{X}}^{\scriptsize\textrm{PBE}}+
E_{\tiny\textbf{C}}^{\scriptsize\textrm{PBE}}.
\]
Another one is based on a screened Coulomb potential for the exchange 
interaction.\cite{Heyd1,Heyd2,Paier} Among all the other Coulomb 
terms of the Hamiltonian, only the exchange interaction uses a 
screened Coulomb potential. The exchange component of the XC energy 
is divided into two parts as long-range (LR) and short-range (SR) while 
the correlation component is represented by standard PBE~\cite{pbe} 
functional. In this hybrid approximation, which reduces the 
self-interaction (SI) error of the DFT, the exchange energy is given as,
\[
E_{\tiny\textbf{X}}^{\scriptsize\textrm{HSE}}=
a E_{\tiny\textbf{X}} ^{\scriptsize\textrm{HF,SR}}(\omega)+
(1-a)E_{\tiny\textbf{X}} ^{\scriptsize\textrm{PBE,SR}}(\omega)+
E_{\tiny\textbf{X}} ^{\scriptsize\textrm{PBE,LR}}(\omega)
\]
where $a$ is the mixing coefficient that is determined by perturbation 
theory~\cite{pbe2} and $\omega$ is the range separation parameter 
controlling the extent of the SR interactions.\cite{Heyd1,Heyd2,Paier}
We performed Heyd-Scuseria-Ernzerhof hybrid functional 
(HSE06)~\cite{Heyd1,Heyd2,Paier} calculations with the Vienna 
Ab-initio Simulation Package (VASP)~\cite{vasp}. We used projector
augmented waves (PAW)\cite{paw1,paw2} datasets and optimized
atomic positions by minimizing the forces requiring a precision of
0.01 eV/{\AA}.  For the computation of the absorption spectra,  the 
transitions from occupied to unoccupied states are considered within 
the first Brillouin zone. The imaginary part of the dielectric 
function $\varepsilon_2(\omega)$ can then be determined by 
the summation,

\begin{eqnarray*} 
\varepsilon^{(2)}_{\alpha \beta}(\omega)=\frac{4 \pi^2 e^2}{\Omega}
\lim_{q\to 0}\frac{1}{q^2}&&\sum_{c,v,\mathbf{k}}
2\textsl{w}_{\mathbf{k}}\delta(\epsilon_{c\mathbf{k}}-\epsilon_{v\mathbf{k}}-\omega) \\
&& \hspace{-2mm}\times\langle u_{c\mathbf{k}+\mathbf{e}_\alpha q} 
\vert u_{v\mathbf{k}} \rangle\langle u_{c\mathbf{k}+\mathbf{e}_\beta q} 
\vert u_{v\mathbf{k}} \rangle^*
\end{eqnarray*}

\noindent where the indices $c$ and $v$ denote empty and 
filled levels respectively, $ u_{c\mathbf{k}}$ are the cell periodic 
part of the orbitals and $\textsl{w}_{\mathbf{k}}$ are the weight factors at each 
$k$-point $\bf k$.\cite{Gajdos} 

In order to study the effect of the solvent environment (chloroform) 
on the electronic structure of the dye molecules, we performed 
calculations using the new non-linear polarizable continuum 
model (PCM) and its linear  counterpart, both of which has been 
implemented in the open-source code JDFTx.\cite{jdftx,Gunceler} 
These PCM approaches enable us to make better comparisons with 
experiment.

\begin{figure*}[t!]
\epsfig{file=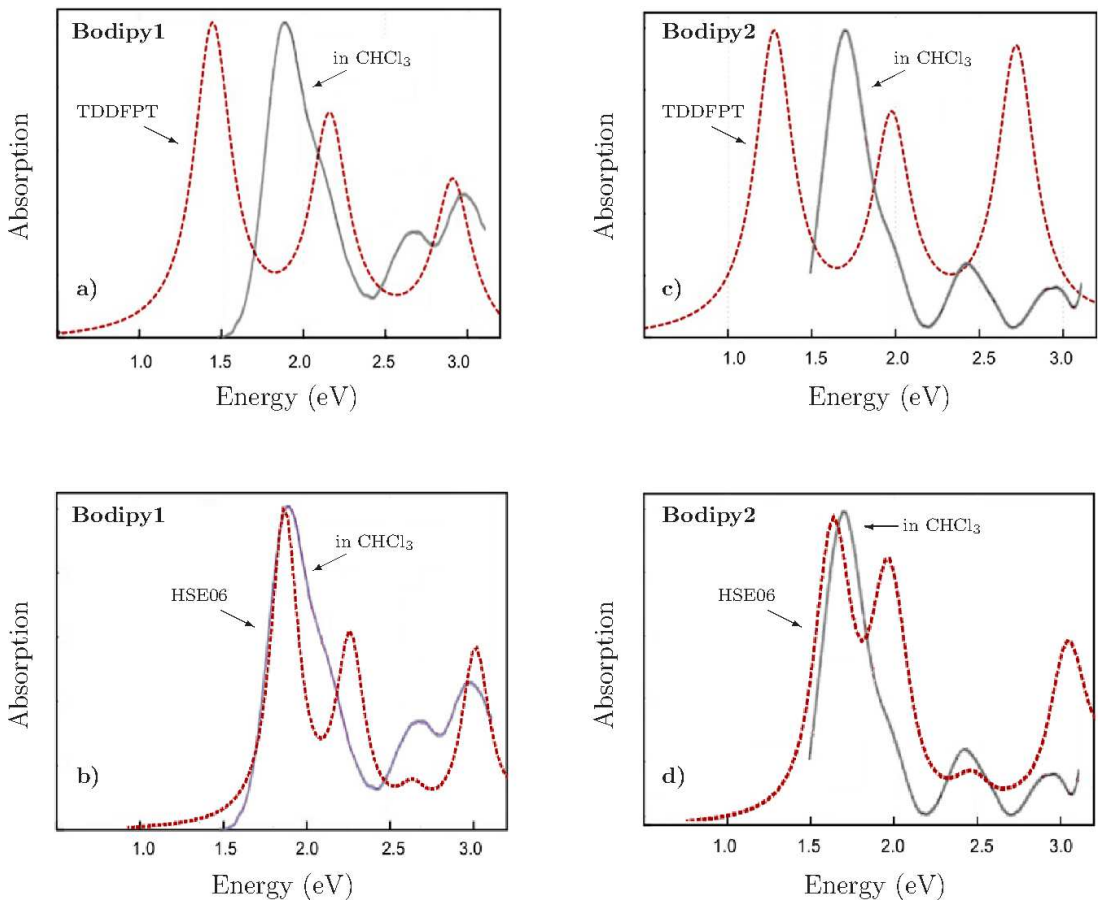,width=14.5cm}
\caption{Schematic comparison of the calculated and the experimental optical spectra for 
Bodipy dye sensitizers. Experimental spectra are taken from Ref.~\cite{Kolemen}\label{fig2}}
\end{figure*} 

\section{Results \& Discussion}

The placement of substituents that conjugated to the Bodipy core (Fig.~\ref{fig1}) 
is very important for efficient charge injection.  The absorption peaks of the 
compounds can be shifted to the lower energies, the intensity and broadening 
of those peaks rise depending on the increment of the added arms. We know 
that the parent Bodipy dye shows weak absorbance at only shorter 
wavelengths.\cite{Kumaresan} Addition of the styryl arms intensifies 
and shifts this major peak to the longer wavelengths while producing new 
weak absorbances at higher energies.\cite{Ela,Kumaresan}

In our case there are four absorption peaks of each compound 
(Fig.~\ref{fig2}). The Bodipy1 and Bodipy2 compounds absorb 
strongly at around 699 nm and 746 nm, respectively, in CHCl$_3$.\cite{Ela,Kolemen}. 
These lowest lying excitations involve $S_0\rightarrow S_1$ transitions 
which asymmetrically  increase the charge density on the meso-carbon
and decrease it in the other parts.\cite{Ela} Such a directional charge 
redistribution is hoped to achieve better charge injection. TDDFPT predicts 
three distinct absorption peaks which are red-shifted similarly for both 
Bodipy1 and Bodipy2, while the width of their lowest 
lying excitations are well-described. 

\begin{table*}
\caption{Maximum optical absorption wavelengths, $\lambda_{\scriptsize\textrm{max}}$ (nm), 
and HOMO-LUMO energy gaps, E$_{\scriptsize\textrm{gap}}$ (eV), of compounds Bodipy1 and 
Bodipy2.\label{table1}}
\vspace{3mm}\small
\begin{tabular*}{\textwidth}{@{\extracolsep{\stretch{1}}}*{9}{c}@{}} 
&&Exp & PBE & TDDFPT & HSE & PBE0 & HSE+PCM$^a$ & PBE0+PCM$^a$\\[1mm] \hline
\multirow{2}{*}{Bodipy1} & E$_{\textrm{\scriptsize gap}}$ & 1.57 & 0.60 & 1.21 & 1.58 & 2.15 & 1.61 & 2.17\\
&$\lambda_{\scriptsize \textrm{max}}$ & 699$^b$ & 1542 & 867 & 695 & 450 & 684 & 522 \\ \hline
\multirow{2}{*}{Bodipy2} & E$_{\textrm{\scriptsize gap}}$ & 1.43 & 0.90 & 0.96 & 1.39 & 1.95 & 1.41 & 1.97\\
&$\lambda_{\scriptsize \textrm{max}}$ & 746$^b$ & 1040 & 960 & 764 & 568 & 755 & 563\\ \hline 
\end{tabular*}
\begin{flushleft}
$^{\rm a}$ Nonlinear PCM included for CHCl$_3$ \\
$^{\rm b}$ Absorption data collected in CHCl$_3$ solution, Ref.~\cite{Kolemen}
\end{flushleft}
\end{table*}

The HOMO-LUMO energy level characterization of the compounds had 
been made by cyclic voltammetry experiments~\cite{Ela,Kolemen}. 
Standard DFT calculations underestimate the HOMO-LUMO gaps 
because of the SI error in the Coulomb term (Table~\ref{table1}). 
Interestingly, the prediction of the gap with PBE for Bodipy1 of 
0.60 eV is much worse than that of Bodipy2 being 0.90 eV. We also 
note that the correction to the energy gap by TDDFPT is much better for 
Bodipy1. Time dependent density functional perturbation approach 
shifts PBE-predicted absorption peaks toward longer wavelengths 
in the near IR region (Fig.~\ref{fig2}). Although TDDFPT improves over 
the standard DFT gaps, there still remains a noticeable underestimation 
due to low-lying excitation energies associated with significant charge 
transfer (CT) errors. The extent of the error depends on the spatial 
separation of charge excitation from one carbonyl group to anot
one.\cite{Peach} Other than the presence of CT states, the determination 
of optical spectrum of such a large molecular system over a broad 
frequency range is computationally expensive with linear response 
TDDFT. This has been made possible by TDDFPT with drastically less 
computational effort, but the main drawback that still exists is the 
localization of XC functionals, which yields the underestimation 
in orbital energy differences~\cite{Malcioglu,Dreuw,Dreuw2,Peach}. 

Because of the necessity of non-local XC functionals to fix this 
problem, hybrid approaches with semi-local exchange terms have 
been used before.\cite{Dreuw,Peach} Although hybrid functionals do 
not include the excitation process, some CT states, depending on 
spatial orbital overlaps, can be generated with these functionals.\cite{Peach} 
In our case here, for Bodipy1 and Bodipy2 molecules,  the occupied 
and virtual orbitals do not overlap. The CT excitation energies as 
mean orbital energy differences improve reasonably by using HSE06 
hybrid XC functional which incorporates partial exact exchange 
energy through a screened Coulomb interaction. HSE06 method 
predicts HOMO-LUMO gaps as 1.58 eV and 1.39 eV for Bodipy1 and 
Bodipy2, respectively (Table~\ref{table1}). These values show good 
agreement with the experimental counterparts, as also reflected 
in Fig.~\ref{fig2}b and Fig~\ref{fig2}d,  from the corresponding 
optical absorption thresholds involving excitations between the 
frontier molecular orbitals. For both of the dye molecules, HSE06 
calculations also achieve a reasonable estimation the widths of 
those lowest lying peaks. Moreover, the positions of higher lying 
absorption peaks are properly described.

The PBE0 hybrid XC functional, on the other hand, significantly 
overestimates HOMO-LUMO gaps for both Bodipy1 and Bodipy2. 
Corresponding maximal absorption peak positions are shifted to 
shorter  wavelengths in the visible region. The directional
charge redistribution upon a light harvesting excitation on 
these dye molecules indicates the need for a proper description
of the localization effects. In HSE06 functional, the mixing 
parameter and range of separation can be considered as an 
effective screening of the Coulomb interaction. In this way, 
a reasonable description of such localized states can practically 
be obtained. The main difference of HSE06 from PBE0 is 
the omission of long range part of nonlocal exact exchange 
and compensation of it with semilocal PBE exchange.
For extended states and materials this may not cause much
of a change in the calculated properties. Here, for Bodipy 
molecules the screened exact exchange kernel acts in favor
of the HSE06 approach.   PBE0 functional tend to substantially
overestimate  the gaps for Bodipy dyes due to relatively stronger 
charge localization as a result of the insufficient screening of 
the Coulomb interaction. Our PBE0 results are in agreement with 
previous theoretical studies such that the existence of the
long-range HF exchange strongly enlarges the band gaps.\cite{Gerber}

Absorption spectra measurements reported in Ref.~\cite{Kolemen} 
were collected in CHCl$_3$ solution, the effect of which was 
ignored in our previous calculations. To quantify the effect of solvent, 
we calculated HOMO-LUMO gaps of Bodipy derivatives embedded 
in a nonlinear polarizable continuum environment~\cite{Gunceler}.
We obtained a HOMO-LUMO gap enhancement of about 0.2 eV
with PBE0 and HSE06 for both dye molecules (Table~\ref{table1}).
The difference from vacuum results is small because CHCl$_3$, 
with a bulk dielectric constant of 4.81, is a weakly polar solvent
and does not interact strongly with the dye. For the same reason, 
dielectric saturation effects are not pronounced and same results 
can be obtained with a linear polarizable continuum model.
For ionic electrolytes in DSSC applications, solvent effects might 
become much more important.

\section{Conclusions}
Our calculations revealed that Bodipy derivatives can be used as 
good sensitizers exhibiting strong absorption characteristics
toward the longer wavelengths in the visible and near-IR region.
We have performed calculations with TDDFPT as well as HSE06 and 
PBE0 hybrid functionals to determine electronic and  optical properties 
of candidate dye molecules for DSSC applications. Lowest lying excitations
in Bodipy-based sensitizers exhibit directional charge redistribution with 
no overlap between the occupied and virtual orbitals. When standard 
XC functionals are used TDDFPT calculations can not generate 
excitation energies of CT states accurately. PBE0 hybrid functional
largely overestimates HOMO-LUMO gaps due to the presence of 
long-range exact exchange giving rise to strong localization effects. 
Screened Coulomb HSE06 functional reasonably describes  the gaps 
and photo absorption characteristics indicating the effect of screening
in the exchange term. We have shown that solvation models have a 
nonnegligible contribution in improving the band gaps of Bodipy 
dyes even in non polar solutions with low dielectric constants.
Nonlinear PCM effects on the absorption characteristics of dye sensitizers 
might become more pronounced for polar electrolytes.

\begin{acknowledgments}
Financial supported is acknowledged from T\"{U}B\.{I}TAK, 
The Scientific and Technological Research Council of Turkey 
(Grant \#110T394). Computations were performed at 
ULAKB\.{I}M, Turkish Academic Network and Information Center.
\end{acknowledgments}

\end{document}